\documentclass[11pt]{article}

\usepackage[margin=1in]{geometry}
\usepackage{amsmath,amssymb,amsthm}
\usepackage{booktabs}
\usepackage{tabularx}
\usepackage{array}
\usepackage{enumitem}
\usepackage{microtype}
\usepackage{xcolor}
\usepackage[hidelinks]{hyperref}

\newcolumntype{Y}{>{\raggedright\arraybackslash}X}
\newcolumntype{P}[1]{>{\raggedright\arraybackslash}p{#1}}

\newtheorem{proposition}{Proposition}
\newtheorem{definition}{Definition}

\newtheorem{counterexample}{Counterexample}

\newcommand{\project}{\mathsf{project}}
\newcommand{\fork}{\mathsf{fork}}
\newcommand{\settle}{\mathsf{settle}}

\newcommand{\prefix}{\mathsf{prefix}}

\newcommand{\code}[1]{\texttt{#1}}

\title{Replay Is a Family of Assertions:\\
Fork-Safety in Event-Sourced Agent Runtimes}
\author{Yohei Nakajima\\Untapped Capital / activegraph.ai}
\date{August 2026}

\begin{document}

\maketitle

\begin{abstract}
Event-sourced agent runtimes commonly claim deterministic replay, cheap
forking, and lineage.  These are families of assertions, not one guarantee.  A
fixed-log fold may be deterministic while reactive settlement remains
schedule-sensitive; a recorded oracle result can enable zero-call replay while
the original interaction still changed cost or world state; and a prefix copy
cannot undo an action in a discarded suffix.

We present an executable F\# contract that separates facts from effect
requests, makes scheduling explicit, and assesses a cut relative to projection,
strict execution replay, external continuation, or a declared
counterfactual-world observation.  Preserved counterexamples show that general
settlement is neither terminating nor confluent and that disjoint declared
writes do not prevent read/trigger interference.  Against 5,540 public runtime
logs, all 317,296 projection cuts satisfy the stated observation, while 36,251
cuts split a model request and do not license continuation; another 120,969
are Conditional because a committed oracle result must be served from the
record without re-execution.  In 121 observed forks, 16,625 retained-prefix
events have zero structural mismatches, but none crosses an external-effect
boundary.  A separate public offline conformance fork crosses a committed,
recorded oracle boundary.  Its generator observes that the inherited result is
served without a second fixture-oracle call; its verifier checks receipt/log
consistency and a fork-bound HMAC assertion under a public conformance key.  The
result reframes cheap forking as a property-indexed semantic contract with
explicit evidence and environment preconditions.
\end{abstract}

\noindent\textbf{Artifact status.}
The code, checked-property suite, conformance vectors, and evidence ledger
accompany this manuscript.  The Synthetic Players capsule and the public
activegraph-bridge v0.2.0 post-oracle fixture are revision- and hash-pinned.

\section{Introduction}

Agent runtimes increasingly retain an append-only record of model calls, tool
results, human decisions, state updates, and coordination events.  Event
sourcing makes several desirable properties seem immediate.  If state is a
fold over the log, a run can be replayed.  If a child log shares a prefix with a
parent, a run can be forked without recomputing that prefix.  If causes are
recorded, artifacts can be traced back to model and tool activity.  ActiveGraph
states these properties as consequences of making the log the source of truth
\cite{nakajima2026log}.

This paper formalizes and empirically checks the replay and fork properties
claimed on architectural grounds in Nakajima (2026).  The relationship is
deliberate: the earlier work describes a deployed architecture, while the
present work asks which side conditions make its architectural claims sound.
Negative results about the originating runtime are therefore part of the
contribution rather than exceptions to be hidden.

Two gaps motivate the work.  First, ``deterministic replay'' often means
deterministic replay \emph{after oracle outputs have been fixed}.  This is
useful, but the phrase suppresses operational questions.  Was the model output
captured faithfully?  Is a request retained without its result?  Does replay
serve the recorded result or make a new call?  Is the runtime itself
schedule-independent, or does a particular FIFO order merely produce one
repeatable execution?

Second, event sourcing in a business application normally reconstructs
application state; it does not claim to reverse the external world.  That
distinction becomes load-bearing for agents.  A retrospectively discarded log
suffix may contain an email, a charge, a delete, a publication, or a billable
model call.  The forked trace can omit the effect, but the world cannot be
assumed to be one in which the effect never happened.  A correct fork contract
must say which property it preserves.

Figure~\ref{fig:replay-family} names seven assertions commonly compressed into
the word ``replay.''  They are ordered here from artifact identity toward
claims about execution and the external world, but they do not form a simple
implication chain.  A system can preserve a decision while changing its path,
or reproduce bytes while retaining too little in-log evidence to execute.

\begin{figure}[ht]
\centering
\small
\begin{tabularx}{0.96\textwidth}{@{}>{\bfseries}p{0.22\textwidth}Y@{}}
\toprule
Assertion & Question answered \\
\midrule
Byte & Is the serialized artifact identical under its verifier? \\
Trace & Are the same ordered semantic events retained? \\
Projection & Does folding those events yield equivalent selected state? \\
Path & Is the model, tool, and action sequence the same? \\
Decision & Is the task-level conclusion equivalent? \\
Execution & Can code re-execute with zero live calls by serving recorded results? \\
Environment & Does the observed external world match the world claimed by the prefix? \\
\bottomrule
\end{tabularx}
\caption{Replay is a family of assertions.  Each row requires its own
observation and evidence contract.}
\label{fig:replay-family}
\end{figure}

Table~\ref{tab:assertion-map} makes the scope of those seven assertions
explicit.  The four fork properties operationalize only the assertions needed
to decide whether a particular cut may be projected, executed, continued, or
treated as a counterfactual world.  Path and Decision remain application-level
equivalences rather than independent fork properties in this artifact.

\begin{table}[ht]
\centering
\small
\begin{tabularx}{\textwidth}{@{}P{0.14\textwidth}P{0.34\textwidth}Y@{}}
\toprule
Assertion & Operationalization in this paper & Evidence boundary \\
\midrule
Byte & Source-artifact verifier; not a fork property & External capsule verification can establish byte identity without raising a run's log-derived grade. \\
Trace & ExactTrace and the checked prefix identities & The observed-fork audit compares semantic prefix fields modulo declared lineage exclusions. \\
Projection & ProjectionReplay under ProjectedState or FactsOnly$(K)$ & A fixed, well-formed ordered prefix is sufficient; no effect execution occurs. \\
Path & ExactTrace restricted to model, tool, and action events when a source schema exposes them & Not independently assessed in the submitted cut matrix; equal decisions need not imply equal paths. \\
Decision & A source-specific task-level equivalence when decisions are present & Not an independent fork property or replay-grade license in this artifact. \\
Execution & StrictExecutionReplay & Boundary is the first run-level grade that licenses recorded-effect re-execution in general; the cut-level assessment remains authoritative. \\
Environment & ExternalContinuation and CounterfactualWorld$(\Omega)$ & No replay grade suffices alone; footprint coverage, a target-environment premise, and any reconciliation obligations are also required. \\
\bottomrule
\end{tabularx}
\caption{Cross-map from the replay assertion family to the paper's formal and
empirical scope.  Replay grades summarize retained evidence; they are not a
one-to-one replacement for these assertions.}
\label{tab:assertion-map}
\end{table}

The familiar reducer shape

\begin{equation}
  A : S \times E \rightarrow S \times F^*
\end{equation}

is not the contribution.  It is a Mealy-style transition system
\cite{mealy1955method}, resembles the Elm update architecture
\cite{elmarchitecture}, and appears in functional event sourcing through the
Decider pattern \cite{chassaing2021decider}.  The operational seam needed here
is a scheduled step over configurations plus an interpreter outcome tied back
to a request identity:

\begin{equation}
  \mathsf{step}:C\times\Sigma\rightarrow C,
  \qquad
  \mathsf{interpret}:R\rightarrow
  \{\mathsf{Committed},\mathsf{Failed},\mathsf{Unknown}\}.
\end{equation}

The second function is outside the pure kernel.  Evidence links its result to
the request in the log.  This makes scheduling, incomplete boundaries, and
already-realized external effects explicit without claiming a new minimal
agent calculus.

Table~\ref{tab:prior-claims} states the relationship to the predecessor paper
directly.  The earlier architecture remains useful; the present paper supplies
the contract under which its claims are licensed.

\begin{table}[ht]
\centering
\small
\begin{tabularx}{\textwidth}{@{}P{0.18\textwidth}P{0.20\textwidth}Y@{}}
\toprule
Claim in \cite{nakajima2026log} & Status here & Qualification \\
\midrule
Deterministic replay & Upheld, qualified & A fixed ordered log projects deterministically.  Schedule-independent log production and zero-call execution require separate conditions (Sections~\ref{sec:model}--\ref{sec:fork}). \\
Cheap forking & Upheld, qualified & Prefix copying is structurally cheap; continuation and external-world claims are property-relative (Sections~\ref{sec:fork} and~\ref{sec:validation}). \\
End-to-end lineage & Upheld in the observed scope & The 121 observed children preserve compared prefix fields.  Stronger replay licenses require request/outcome and provenance evidence (Sections~\ref{sec:grades} and~\ref{sec:validation}). \\
\bottomrule
\end{tabularx}
\caption{Claim status relative to \emph{The Log is the Agent}.}
\label{tab:prior-claims}
\end{table}

\paragraph{Contributions.}
This manuscript makes four contributions.

\begin{enumerate}[leftmargin=*]
  \item It defines property-relative fork assessment for projection, strict
  execution replay, external continuation, and a declared external-world
  observation.

  \item It separates effect evidence along external footprint, replay source,
  and lifecycle, allowing a recorded result and an irreversible footprint to
  coexist in one descriptor.

  \item It gives an executable scheduler and preserves counterexamples showing
  that settlement need not terminate or converge.  Disjoint declared writes
  remain insufficient when reads and emitted triggers interfere.

  \item It checks the contracts against 5,540 public runtime logs, every cut in
  those logs, and 121 observed forks.  A public offline conformance fixture
  additionally crosses a committed recorded oracle boundary: its generator
  observes no second fixture-oracle call, and its verifier checks a fork-bound
  receipt and caller assertion.
\end{enumerate}

\paragraph{Genre.}
This is not a mechanized-proof paper.  Model identities are stated as
propositions; empirical claims are encoded as properties, checked over
generated families with FsCheck, and evaluated exhaustively over the cuts of
the available logs.  We call these executable contracts or checked properties
in the running text.  A failed property produces a minimized, named
regression fixture.  Passing checks are finite evidence over the specified
generators and artifacts, not universal proofs.  This choice is intentional:
calculi with
machine-checked composition, termination, and information-flow results already
exist \cite{liu2026lambdaa,garby2026llmbda}.  Our comparative advantage is an
executable runtime contract connected to real runtime artifacts.

\paragraph{Roadmap.}
Section~\ref{sec:model} defines the executable model and observations.
Section~\ref{sec:confluence} reports termination and confluence results.
Section~\ref{sec:fork} defines fork properties and effect conditions.
Section~\ref{sec:grades} connects those conditions to replayability grades.
Sections~\ref{sec:method} and~\ref{sec:validation} describe the artifact and
empirical validation.  We then position the work, discuss implications, and
state limitations.

\section{Executable Runtime Model}
\label{sec:model}

The model is intentionally small, but it is not a claim of expressive
minimality.  Its purpose is to make replay, scheduling, and effects observable
enough to falsify claimed properties.

\subsection{Events, projection, and identity}

An event envelope is a tuple

\begin{equation}
 e = (id, run, seq, kind, cause, emitter).
 \label{eq:event-envelope}
\end{equation}

The event kind is a closed sum containing domain facts, signals, effect
lifecycle events, and evidence facts.  The evidence branch includes
\code{HazardDetected(detail)}; an envelope records it as
\code{EvidenceRecorded(HazardDetected detail)}, and the grader treats that fact
as an explicit blocker.  A log
$L = \langle e_1,\ldots,e_n\rangle$ is ordered and append-only.  The pure
reducer $\mathsf{evolve}:S\times E\rightarrow S$ yields

\begin{equation}
 \project(L) = \mathsf{fold}(\mathsf{evolve}, S_0, L).
 \label{eq:project}
\end{equation}

Event identity is idempotent at the reducer boundary: if an event identifier
has already been applied, a repeated envelope does not mutate state again.  A
duplicate identifier remains an integrity fault for fork assessment, because
idempotent reduction must not silently certify a malformed trace.

State contains domain facts and signals, projected effect descriptors,
evidence facts, integrity faults, and the set of applied event identifiers.
When comparing projected domain state, the implementation excludes the
bookkeeping set of applied identifiers.  This prevents a replay that merely
renames envelopes from being declared semantically different while still
allowing exact-trace equality to detect the change.

\begin{proposition}[Closed-log projection determinism]
For a fixed ordered event log and a pure reducer, repeated evaluation of
Equation~\ref{eq:project} returns the same state.  Oracle calls are not an
exception: their \emph{recorded results}, when present, are events in the fixed
log.
\end{proposition}

This proposition is the trivial base case.  It says nothing about how the log
was produced, whether the event order was unique, whether missing oracle
results can be recovered, or whether effects may be executed again.

\subsection{Behaviors and actions}

A behavior is a tuple

\begin{equation}
 b=(id_b, trigger_b, fire_b, W_b)
\end{equation}

where $trigger_b:S\times E\rightarrow\mathbb{B}$ decides whether an event
enables the behavior, $fire_b:S\times E\rightarrow A^*$ returns actions, and
$W_b$ is a declared write set used only as a diagnostic.  Actions can emit a
domain fact or signal, or request an external effect.  Behaviors cannot emit an
effect outcome or evidence attestation directly; those facts must cross the
interpreter or validation boundary.  This prevents a reducer from minting its
own proof that an external action succeeded or that a run was verified.

The kernel does not execute actions.  Emissions become pending events.  Effect
requests become pending request events and later block until an interpreter
injects a committed, failed, or unknown outcome.  Ordinary .NET tasks, network
clients, mutable caches, and database adapters may exist outside the boundary;
none are reachable from $trigger$ or $fire$.

\subsection{Configurations and scheduling}

A configuration contains projected state, pending events, enabled activations,
outstanding effects, the retained trace, the behavior firing trace, and the next
sequence and generated-identity counters.  A scheduler has two explicit
selectors: one chooses a pending event, and the other chooses an enabled
activation.  Four policies are included: canonical, reverse-event,
reverse-activation, and both reversed.

Writing a configuration as $\langle s,Q,A,O,L_{\mathrm{ret}},T\rangle$, the two load-bearing
transition rules are shown in Figure~\ref{fig:small-step}.  The helper
$\mathsf{materialize}$ turns a behavior's pure action descriptions into pending
fact/signal events or effect-request events.  It never produces an interpreter
outcome.

\begin{figure}[ht]
\centering
\small
\begin{minipage}{0.96\textwidth}
\[
\frac{
 e=\mathsf{selectEvent}_{\sigma}(Q)\quad
 s'=\mathsf{evolve}(s,e)\quad
 A'=A\mathbin{+}\{(b,e)\mid \mathsf{trigger}_b(s',e)\}
}{
 \langle s,Q,A,O,L_{\mathrm{ret}},T\rangle
 \rightarrow_{\sigma}
 \langle s',Q\setminus e,A',O,L_{\mathrm{ret}}\mathbin{+}e,T\rangle
}
\quad\textsc{Process-Event}
\]
\[
\frac{
 a=(b,e)=\mathsf{selectActivation}_{\sigma}(A)\quad
 X=\mathsf{fire}_b(s,e)\quad
 (Q',O')=\mathsf{materialize}(Q,O,X)
}{
 \langle s,Q,A,O,L_{\mathrm{ret}},T\rangle
 \rightarrow_{\sigma}
 \langle s,Q',A\setminus a,O',L_{\mathrm{ret}},T\mathbin{+}(b,e)\rangle
}
\quad\textsc{Fire-Activation}
\]
\end{minipage}
\caption{Small-step scheduling rules.  Event and activation selection are
explicit.  Emissions become visible to state only after a later
\textsc{Process-Event} step, not within the activation batch for one trigger.}
\label{fig:small-step}
\end{figure}

Processing a pending event first applies it to state and then computes enabled
behaviors from the updated state.  Enabled behaviors fire sequentially, but
their emissions enter the pending queue.  Activations handling the same
triggering event therefore observe the same projected state.  A behavior
activated by a later-processed event observes changes committed by earlier
\textsc{Process-Event} steps, including events materialized from prior
activations.  This cross-event ordering seam mirrors the one found in
ActiveGraph and is sufficient for the read/trigger interference witness in
Section~\ref{sec:confluence}.

Let $C\rightarrow_\sigma C'$ denote one legal event-processing or
activation-firing step under scheduler $\sigma$.  Settlement repeatedly steps
until one of three outcomes is reached:

\begin{itemize}[leftmargin=*]
  \item \textbf{Quiescent}: no pending events, enabled activations, or
  outstanding effects;
  \item \textbf{BlockedAwaitingEffect}: no internal work remains, but at least
  one request has no terminal outcome;
  \item \textbf{Diverged}: a caller-supplied step bound is reached before either
  terminal outcome.
\end{itemize}

The step bound is checked only after testing for a terminal configuration.  A
run that becomes quiescent on the exact final step is therefore not
misclassified as divergence.  Conversely, reaching the bound does not prove a
cycle; the kernel reports \code{StepBoundReached} and makes no stronger claim.
BlockedAwaitingEffect is terminal for the settlement procedure but is not a
quiescent normal form.  The confluence definition in Section~\ref{sec:confluence}
ranges only over executions that reach Quiescent; uniqueness under blocking is
a different property and is not claimed here.

\subsection{Observations and equivalence}

Confluence and fork claims are meaningless without specifying what is observed.
The artifact implements three equivalences:

\begin{enumerate}[leftmargin=*]
  \item \textbf{ExactTrace}: envelope and event lists are structurally equal;
  \item \textbf{ProjectedState}: domain projection is equal after excluding
  applied-event bookkeeping and envelope metadata;
  \item \textbf{FactsOnly}$(K)$: the selected fact keys agree.
\end{enumerate}

The validation additionally reports byte equality when a source artifact
provides it, decision equality when the study records decisions, and structural
prefix equality modulo explicitly excluded lineage fields.  These relations
must not be substituted for one another.  In particular, equal decisions do
not imply equal paths, and equal projected state does not imply an unchanged
external world.

\section{Confluence and Quiescence}
\label{sec:confluence}

The ActiveGraph architecture describes behaviors reacting until quiescence
without a privileged orchestrator.  The working graph is then described as a
deterministic projection of the log.  Projection of a \emph{fixed} log is
deterministic, but the reactive loop can produce different logs under different
legal schedules.  We separate termination from uniqueness of the terminal
projection.

\begin{definition}[Termination at a bound]
For behavior set $B$, initial configuration $C_0$, scheduler $\sigma$, and
bound $m$, $\settle_m(B,C_0,\sigma)$ terminates if it reaches Quiescent or
BlockedAwaitingEffect in at most $m$ steps.  StepBoundReached is evidence of
bounded nontermination, not a proof of infinite reduction.
\end{definition}

\begin{definition}[Schedule confluence under an observation]
A behavior set is confluent for $C_0$ and observation $P$ when every legal
scheduler execution that reaches a quiescent configuration reaches a unique
normal form modulo $\simeq_P$.  Formally, if
$C_0\rightarrow^*_{\sigma_1} C_1$ and
$C_0\rightarrow^*_{\sigma_2} C_2$, and both $C_1$ and $C_2$ are quiescent,
then $C_1\simeq_P C_2$.
\end{definition}

This proposition is structurally analogous to the Church--Rosser property, but
the current artifact does not prove a Church--Rosser theorem.  It searches scheduler policies, generates restricted
families, and preserves witnesses when the proposition fails.  Newman's lemma
connects termination plus local confluence to confluence in an abstract
rewriting system \cite{newman1942}; our first counterexample removes the
termination premise before the second examines uniqueness of quiescent forms.

\subsection{A1: general termination is false}

\begin{counterexample}[Self-triggering emission]
Let a behavior trigger on signal \code{loop} and emit the same signal.  Seed the
pending queue with \code{loop}.  Every processed signal enables the behavior,
and every firing enqueues another signal.  No quiescent or effect-blocked
configuration is reached.  For every generated bound from 1 through 100, the
artifact reports StepBoundReached at that bound.
\end{counterexample}

The example is intentionally small.  It shows that purity of the reducer and
append-only logging do not imply termination of a reactive runtime.  Detecting
a repeated finite fingerprint would not solve the general problem: generated
identities and sequence counters can make configurations syntactically fresh
while their domain behavior cycles.  The implementation therefore avoids a
heuristic \code{CycleDetected} verdict.

Two natural restoring contracts remain candidates.  One assigns every
internally emitted event a well-founded rank that strictly decreases along
trigger edges.  The other requires an acyclic graph from event kinds to
behavior emissions.  Both would exclude the fixture, but neither is claimed
minimal, and both need refinement for state-dependent triggers.

\subsection{A2: general confluence is false}

\begin{counterexample}[Overlapping writers]
Behaviors $\alpha$ and $\beta$ are both enabled by \code{start}.  Each emits a
write to fact \code{winner}; $\alpha$ writes 1 and $\beta$ writes 2.  Under
canonical activation order the emitted events are processed so that the
terminal winner is 2.  Reversing activation order yields 1.  Both executions
quiesce, but their projected states differ.
\end{counterexample}

FsCheck generalizes this witness over generated integer values by constructing
distinct adjacent values.  Across the generated family, canonical and reverse
activation schedules always disagree.  The write-set diagnostic correctly
reports \code{winner} as a conflict.

It is tempting to restore confluence by requiring declared write sets to be
pairwise disjoint.  That condition works for a deliberately restricted family:
single-trigger behaviors that each unconditionally write a distinct key and do
not read other behaviors' regions or emit cross-triggering signals.  The suite
generates up to eight such writers and checks all four scheduler policies; their
projected states agree.  The structural reason is stronger than those four
samples: every generated key has exactly one unconditional writer, and no
writer reads or triggers another, so any permutation applies the same set of
commuting updates to distinct map entries.  The four policies are executable
regression coverage for that restricted family, not an enumeration of arbitrary
runtime interleavings.

Declared disjointness is not sufficient in general.

\begin{counterexample}[Read/trigger interference with disjoint writes]
Two behaviors, \code{left} and \code{right}, trigger on \code{start} and emit
signals \code{left-ready} and \code{right-ready}.  Their declared write sets are
empty.  A third behavior triggers on \code{left-ready}, reads whether
\code{right-ready} is already in state, and writes the only fact
\code{observed-right}.  All declared write sets are disjoint.  Under canonical
event order the observer records 0; under reverse event order it records 1.
\end{counterexample}

The second witness matters because a write-write conflict detector would
certify it.  The actual dependency crosses four surfaces: an emission, an event
selection, trigger eligibility, and a state read.  Any useful general restoring
condition must constrain at least behavior reads, behavior writes, emitted
event kinds, and the trigger relation.  Under-declared dependencies must either
be rejected or made observable by instrumentation.
Table~\ref{tab:read-trigger-interference} shows the resulting
schedule-dependent path surface by surface.

\begin{table}[ht]
\centering
\small
\begin{tabularx}{0.96\textwidth}{@{}p{0.18\textwidth}Y Y@{}}
\toprule
Surface & Canonical event order & Reversed event order \\
\midrule
Emission & \code{left-ready}, then \code{right-ready} & \code{right-ready}, then \code{left-ready} \\
Trigger & Observer becomes eligible before right is projected & Observer becomes eligible after right is projected \\
Read & \code{right-ready} absent & \code{right-ready} present \\
Write & \code{observed-right}=0 & \code{observed-right}=1 \\
\bottomrule
\end{tabularx}
\caption{Counterexample 3 crosses emission, event selection, trigger
eligibility, and a state read despite disjoint declared write sets.}
\label{tab:read-trigger-interference}
\end{table}

\subsection{Production ordering masks the question}

The ActiveGraph v1.10.0 scheduler uses a FIFO queue, evaluates matching
behaviors in registration order, and dispatches them sequentially.  Work due at
the same delayed tick remains FIFO.  This supplies a reproducible
\emph{ProductionOrder}, and it may be entirely appropriate as an operational
contract.  It is not evidence that alternate legal orders have the same normal
form.

The distinction is practical.  ActiveGraph computes the match set before the
handler loop, so an earlier handler cannot change which behaviors are already
enabled by that trigger.  It does, however, rebuild each handler's view against
the current graph.  An earlier registered handler can emit an event whose
projection is applied synchronously and changes state read by a later handler
of the same original event; behavior matching for the emitted event occurs at a
later FIFO step.  Relation enumeration adds another ordering seam when a
backend query does not specify a canonical order.

The executable kernel deliberately snapshots state across activations enabled
by one trigger, so it does not reproduce this per-invocation production read
timing.  Its tests vary event and activation order directly, and its
read/trigger witness uses distinct later events.  The production seam is a
source-audited limit on generalization, not a property claimed as validated by
the kernel.  A stable FIFO queue can hide either form of dependence in ordinary
replay tests.

\paragraph{Result.}
The strongest supported conclusion is negative and conditional.  General
termination and general confluence are false.  Isolated disjoint writers
converge for the checked family, but declared writes alone do not restore
confluence.  Deriving and validating a general non-interference contract is
future work; this manuscript does not present a minimal side condition.

\section{Property-Relative Fork Safety}
\label{sec:fork}

A trace fork at cut $k$ retains $L_{<k}$ and gives subsequent events a child
run identity.  For a list representation,

\begin{equation}
  \fork(L,k) = \prefix(L,k).
\end{equation}

That definition makes prefix equality cheap, but it does not by itself make
re-execution or external-world claims sound.  We first describe effects, then
define the properties a fork may be asked to preserve.

\subsection{Effects along three dimensions}

A flat partition into Pure, Idempotent, Cached, and OneShot combines questions
that must be answered separately.  A cached model response describes where
replay material comes from; it does not describe whether the original call had
external cost.  We therefore use three orthogonal dimensions
(Table~\ref{tab:dimensions}).

\begin{table}[ht]
\centering
\small
\begin{tabularx}{\textwidth}{@{}>{\raggedright\arraybackslash}p{0.18\textwidth}>{\raggedright\arraybackslash}p{0.27\textwidth}Y@{}}
\toprule
Dimension & Values & Question \\
\midrule
External footprint & Pure, Idempotent, Compensatable, OneShot, Unknown & What did execution do to the external world, and can that footprint be reconciled? \\
Replay source & Deterministic, Recorded, Uncaptured & Can replay reproduce or serve the result without consulting the live oracle? \\
Lifecycle & Requested, Committed, Failed, Unknown & Is the request/outcome boundary complete and trustworthy at this cut? \\
\bottomrule
\end{tabularx}
\caption{Orthogonal effect dimensions.}
\label{tab:dimensions}
\end{table}

Pure effects are internal descriptions with no external interaction.
Idempotent effects can be repeated or overwritten under an explicit key.
Compensatable effects require a separately verified compensation.  OneShot
effects are irreversible or conservatively treated as non-repeatable.  Unknown
fails closed for world-equivalence claims and becomes an explicit condition for
record-served continuation.  In the validation adapter, model calls are conservatively
OneShot with respect to cost and oracle interaction.  Their outputs may
simultaneously be Recorded, allowing result replay without another call.
These dimensions are orthogonal descriptions, not independent decision rules:
a property may depend jointly on more than one dimension.  In particular,
Idempotent constrains repeated world mutation but does not guarantee that a
second call returns the same bytes, timestamp, identifier, or model output.

Lifecycle is projected from events.  A request must precede exactly one
terminal outcome.  Duplicate requests, conflicting descriptors, an outcome
without a request, or a second outcome after terminal state are integrity
faults.  The first valid request and first valid terminal outcome remain the
projected authority; malformed later events do not overwrite them.

\subsection{Four fork properties}

\begin{definition}[ProjectionReplay]
The child reconstructs the same domain projection as an independently folded
copy of the retained, well-formed, ordered prefix.  No execution occurs.
\end{definition}

\begin{definition}[StrictExecutionReplay]
Every retained effect result needed by execution is deterministic or recorded,
and the cut does not retain a Requested or Unknown lifecycle.  Replaying the
prefix causes zero live oracle or tool calls.
\end{definition}

\begin{definition}[ExternalContinuation]
Relative to a caller-supplied premise that the target environment represents
the retained prefix, the child may continue after strict replay.  Inherited
committed OneShot effects must be served from the record and never executed
again.  A committed Recorded effect with Unknown footprint is Conditional: it
may be served without a live call, but no stronger world-state claim is licensed
until its footprint is resolved.  This property deliberately does not assert that a
retrospectively discarded source suffix left the original world unchanged.
\end{definition}

\paragraph{Remark (assessor scope).}
The v0 assessor takes the target-environment premise as an input assumption; it
does not verify a snapshot or fresh-environment attestation.  Its verdict is
relative to the immutable log snapshot supplied at assessment time.  A
production caller must bind the assessment to a source high-water mark and
establish log closure, quiescence, or another explicit freshness policy before
acting on it.  The companion bridge in Section~\ref{sec:validation}
authenticates and fork-binds a caller-issued environment assertion, but does not
independently validate the assertion's contents or rule out a later-arriving
hazard.  The language-neutral assessor remains verifier-agnostic.

\begin{definition}[CounterfactualWorld$(\Omega)$]
Let $\Omega$ be a declared set of external observables, such as delivered
messages, provider charges, published records, or payment-ledger entries.  In
addition to ExternalContinuation, the actual world and a world in which the
discarded suffix did not occur agree under every observation in $\Omega$.
Discarded committed OneShot or unknown effects whose observables intersect
$\Omega$ block the claim; discarded idempotent or compensatable effects whose
observables intersect $\Omega$ make it conditional on reconciliation or
successful compensation.  Effects proven disjoint from $\Omega$ do not affect
this property.
\end{definition}

The kernel does not inspect the external world.  It checks whether retained
effect evidence is sufficient to license CounterfactualWorld$(\Omega)$ under
the stated footprint-coverage assumption.  The current effect descriptor does
not carry per-effect observable labels, so the executable v0 instantiates
$\Omega$ as the entire external surface recognized by the selected adapter
profile.  Every classified non-Pure effect is therefore conservatively in
scope.  The companion bridge receipt can carry explicit observable tags, but
the reported corpus counts do not retroactively use them.  Supporting a
narrower $\Omega$ in the generic assessor requires adapter-supplied tags; the
reported counts are not claims about such a narrower set.  Even a Sound verdict
is a contract-relative license, not an empirical observation that two worlds
are identical.

Closed assessor return values are typeset as \code{Sound}, \code{Conditional},
and \code{Unsound}; lowercase words retain their ordinary predicate meaning.
These definitions make ``sound'' property-indexed.  The same cut may be Sound
for ProjectionReplay, Conditional for ExternalContinuation, and Unsound for
CounterfactualWorld.  Table~\ref{tab:forkconditions} summarizes the executable
assessment.

\begin{table}[ht]
\centering
\small
\begin{tabularx}{\textwidth}{@{}p{0.22\textwidth}Y Y@{}}
\toprule
Property & Retained-prefix conditions & Discarded-suffix conditions \\
\midrule
ProjectionReplay & Whole source log is structurally well formed; ordering authority is fixed & None for projected prefix equality \\
StrictExecutionReplay & No open/unknown lifecycle; every needed result is deterministic or recorded & None \\
ExternalContinuation & Strict replay; retained committed OneShot effects are not re-executed; committed Recorded Unknown footprint is Conditional & Target environment represents the retained prefix as a caller-supplied premise \\
\shortstack[l]{CounterfactualWorld\\$(\Omega)$} & ExternalContinuation conditions; effect classification covers $\Omega$ & No unreconciled footprint intersecting $\Omega$; in-scope OneShot or unknown blocks \\
\bottomrule
\end{tabularx}
\caption{Property-relative fork conditions.}
\label{tab:forkconditions}
\end{table}

\subsection{Checked prefix identities}

The executable suite checks the following trace identities over generated logs.

\begin{proposition}[Identity fork]
$\fork(L,|L|)$ retains the full trace.  Under ExactTrace, the shared prefix is
equal to $L$; child lineage is stored outside that shared list.
\end{proposition}

\begin{proposition}[Nested prefix collapse]
For $j\leq i\leq |L|$, retaining $i$ events and then $j$ events has the same
shared prefix as retaining $j$ events directly.  The nested branch still records
the outer child as its parent, so raw branch records are not identical.
\end{proposition}

\begin{proposition}[Projection commutation]
For a well-formed ordered log,
\begin{equation}
 \project(\fork(L,k)) = \project(\prefix(L,k)).
\end{equation}
This is a trace/projection statement, not an external-world theorem.
\end{proposition}

\begin{proposition}[Suffix irrelevance is property-relative]
The discarded suffix is irrelevant to ProjectionReplay of the retained prefix.
It is not generally irrelevant to CounterfactualWorld.
\end{proposition}

The last proposition corrects an over-broad formulation of cheap forking.  A
committed OneShot event in the discarded suffix is a concrete witness: the
forked log excludes it and projects correctly, but the action remains true in
the world.  Likewise, a discarded request without a trustworthy terminal
outcome is Unsound because the external system may have executed it even though
the log never learned the result.

\subsection{Cuts through effects}

For each log, the validator constructs every prefix state in one forward scan
and every environmental suffix finding in one reverse scan.  It then assesses
all $|L|+1$ cuts.  Table~\ref{tab:cut-through-request} separates the three
boundary regions created by one request/outcome pair.

\begin{table}[ht]
\centering
\small
\begin{tabularx}{0.96\textwidth}{@{}P{0.20\textwidth}P{0.28\textwidth}Y@{}}
\toprule
Cut & Retained prefix & Licensed claim \\
\midrule
$k_0$ before request & No inherited request & Projection is sound; a later source-side effect may still block a retrospective world claim \\
$k_1$ after request, before outcome & Requested lifecycle only & Strict replay and continuation are refused \\
$k_2$ after recorded commit & Request plus terminal recorded result & Strict replay can be sound; continuation requires zero re-execution \\
\bottomrule
\end{tabularx}
\caption{A single request/outcome pair creates three semantically different
fork regions.}
\label{tab:cut-through-request}
\end{table}

\paragraph{Before a request.}
The retained prefix does not inherit the effect.  Projection replay is sound.
If the source later committed a OneShot effect, CounterfactualWorld is unsound
for a retrospective fork because that discarded action already occurred.

\paragraph{Between request and outcome.}
The prefix contains a Requested lifecycle.  StrictExecutionReplay and
ExternalContinuation are unsound: the effect may be running, may have committed
without acknowledgement, or may have failed.  A step bound or missing response
must not be treated as a safe cancellation.

\paragraph{After a committed OneShot.}
The response may be Recorded, making strict replay possible with zero external
calls.  ExternalContinuation is Conditional: the interpreter must serve the
recorded result and enforce zero re-executions.  This is a semantic obligation,
not merely an optimization.

\paragraph{After an uncaptured idempotent result.}
Idempotence alone does not reconstruct the retained result.  Re-executing the
same key may leave the external resource unchanged while returning a different
receipt, timestamp, identifier, or oracle output.  Because ExternalContinuation
in this paper includes StrictExecutionReplay, the v0 verdict remains Unsound.
A separate live-resumption property could permit a repeated call under a
provider-specific outcome-equivalence contract; it is not the property checked
here.

\paragraph{After a reported failure.}
A terminal Failed event does not establish that the external world was
unchanged.  A non-atomic remote operation may partially mutate state before
reporting failure.  The executable assessor therefore makes discarded failed
Idempotent or Compensatable effects Conditional on reconciliation, and treats
discarded failed OneShot or Unknown effects as Unsound.  Only a failed Pure
effect adds no world finding.

Compensation is likewise not retroactive.  Even successful compensation may
leave a temporal interval in which the mutation was observable and could have
triggered webhooks, quotas, or other secondary effects.  A Conditional
Compensatable verdict licenses CounterfactualWorld only for an $\Omega$ and
environment contract that explicitly exclude or reconcile those exposures.

These cases expose why a Decider that returns only domain events cannot express
the full agent-runtime question.  The log needs request/outcome boundaries,
footprint evidence, and replay-source evidence before it can license execution
or world-level claims.

\section{Replayability Grades as Evidence Licenses}
\label{sec:grades}

The activegraph-bridge project introduced five replayability levels from
observed through native.  The names arose from implementation failure modes.
We make their meaning explicit by assigning each grade a set of licensed
questions (Table~\ref{tab:grades}).

\begin{table}[ht]
\centering
\small
\begin{tabularx}{\textwidth}{@{}>{\raggedright\arraybackslash}p{0.16\textwidth}>{\raggedright\arraybackslash}p{0.32\textwidth}Y@{}}
\toprule
Grade & Required evidence & Licensed use \\
\midrule
Observed & Some inspectable trace or result & Inspection and available lineage only \\
Envelope & Faithful invocation envelope and completion, no lossy-envelope fact & Recorded-output playback; fork before invocation \\
Boundary & Envelope; mediated effects; clean reconstruction; complete known effect boundaries; no blockers & Re-execution using recorded effects; forks at effect boundaries \\
Checkpointed & Boundary plus recorded checkpoint & Resume from a restorable snapshot \\
Native & Checkpointed plus native-runtime evidence & Native replay, fork, and structural diff \\
\bottomrule
\end{tabularx}
\caption{Replay grades and their licenses.}
\label{tab:grades}
\end{table}

Let $Q(g)$ be the set of questions licensed by grade $g$.  The current taxonomy
is defined as the chain

\begin{equation}
 Q(Observed) \subset Q(Envelope) \subset Q(Boundary)
 \subset Q(Checkpointed) \subset Q(Native).
\end{equation}

This chain is a definitional summary, not a substantive theorem.  The artifact
checks its ordering mechanics, but the paper's claim is narrower: a grade
summarizes the questions supported by retained evidence, while a cut-level
assessment decides one proposed fork.  A richer taxonomy could contain
incomparable evidence sets.

\subsection{Computation from the log}

The grade is computed from projected evidence and effect completeness.  It is
not accepted as a source label.  In particular, a \code{NativeRuntime} marker
alone does not grant Native: the log must also satisfy checkpoint, boundary,
and envelope prerequisites.  Unknown footprint, uncaptured results, incomplete
or unknown lifecycles, integrity faults, lossy envelopes, unmediated effects,
and explicit hazards are blockers.

The adapter follows the same fail-closed rule.  It grants evidence only for
exact recognized source event types.  An event whose name merely contains
\code{verification} remains an unclassified domain signal.  A verification
event must contain an explicit Boolean verdict.  The bridge-specific Boundary
derivation additionally requires an effects-served count and an explicit null
divergence: the source payload's \code{divergence} field must be present and its
JSON value must be \code{null}, asserting that the bridge verifier detected no
unresolved reconstruction divergence.  This is logged evidence, not an
independent proof of equivalence.  These requirements prevent an event-name
coincidence from raising the grade.

\subsection{Grades are not monotone in time}

It is natural to view grades as achievements that only rise.  That view is
incorrect for evidence predicates.

\begin{counterexample}[Hazard downgrade]
A prefix records a faithful envelope, completion, mediated boundary, clean
reconstruction, and successful verification.  It grades Boundary.  Appending a
\code{EvidenceRecorded(HazardDetected detail)} event adds an explicit blocker,
so the extended log grades Envelope.
\end{counterexample}

The grade of a \emph{fixed log or prefix} licenses questions about that evidence.
It need not be monotone under extension.  This matters for forks: the grade of
the shared prefix can be higher or lower than the grade of the completed source
run, and a later hazard must not retroactively be ignored.

\subsection{Connection to fork safety}

The grade is a compact mechanism for checking some fork preconditions, but it
does not replace property-relative assessment.  Boundary evidence suggests
that effect results and reconstruction are available, licensing forks at known
effect boundaries.  A specific cut can still split a request from its outcome.
Likewise, a Boundary run can contain a committed OneShot effect that makes a
CounterfactualWorld claim conditional or unsound.

The validation therefore reports both the run-level grade and all cut-level
verdicts.  A grade answers ``what kinds of questions may this evidence support
in principle?''  A fork assessment answers ``does this particular cut satisfy
the conditions for property $P$?''

\section{Executable Specification and Method}
\label{sec:method}

The artifact is a single .NET solution written in F\#.  Purity is enforced at
the transition boundary, not imposed on the surrounding program.  The source
contains ordinary file I/O, JSON parsing, hashing, mutable library internals,
and command-line orchestration outside the reducer.  There is no database,
dependency-injection container, actor framework, plugin system, network client,
or asynchronous interpreter inside the kernel.

\subsection{Artifact structure}

The artifact includes:

\begin{itemize}[leftmargin=*]
  \item a closed event and effect model;
  \item a pure projector and explicit settlement scheduler;
  \item branch construction with parent/child identity, cut, shared prefix,
  continuation, inherited effects, and prefix hash;
  \item cut-level fork assessments and replay-grade computation;
  \item an ActiveGraph/bridge JSONL normalizer and batch validator;
  \item 50 passing tests, including FsCheck properties and named fixtures;
  \item seven language-neutral JSON conformance vectors and a JSON Schema;
  \item read-only SQLite export and actual-fork audit scripts;
  \item a findings log that records failed checked properties as they are
  discovered.
\end{itemize}

All F\# projects treat warnings as errors.  Pattern matches over closed domain
unions are exhaustive.  External effect execution is not linked into the
kernel, so replay cannot accidentally call a model or tool through an injected
client.

\subsection{Property-based checks}

FsCheck generates bounded event values, writer families, effect dimensions,
and step bounds.  Scheduler policies are enumerated rather than sampled because
the executable model exposes four intended extremes.  Properties include:

\begin{itemize}[leftmargin=*]
  \item deterministic projection and event-identity idempotence;
  \item exact-bound quiescence and bounded self-cycle divergence;
  \item convergence of isolated disjoint writers under all schedulers;
  \item schedule dependence for generated conflicting writers;
  \item identity, nested-collapse, projection, and suffix identities for forks;
  \item verdict matrices across generated footprint and replay-source values;
  \item grade-chain mechanics and prerequisite enforcement.
\end{itemize}

The 50-test suite comprises 15 live FsCheck properties and 35 xUnit facts.
Per-property \code{MaxTest} values range from 100 to 250, totaling 3,300
top-level generated trials in a default full run; properties that enumerate
schedulers or dimension values perform additional checks inside each trial.
No fixed global replay seed is configured.  Regression reproducibility comes
from the checked-in named fixtures rather than from repeating one pseudorandom
stream.

FsCheck shrinking is useful for scalar values, but the semantically important
counterexamples are named and checked in as JSON-backed fixtures.  The fixtures
include overlapping writers, read/trigger interference despite disjoint writes,
a self-cycle, a discarded OneShot effect, an unresolved request, malformed
effect transitions, and forged look-alike evidence types.

Table~\ref{tab:vector-coverage} states the deliberately bounded role of the
seven language-neutral vectors.  They are portability smoke tests, not an
exhaustive cross-product of property, footprint, replay source, lifecycle, and
obligation.

\begin{table}[ht]
\centering
\scriptsize
\begin{tabularx}{\textwidth}{@{}P{0.38\textwidth}Y@{}}
\toprule
Vector & Boundary exercised \\
\midrule
\path{projection-last-write-wins} & ProjectionReplay \code{Sound}; deterministic fact projection \\
\path{discarded-one-shot-is-not-counterfactual} & CounterfactualWorld \code{Unsound}; discarded committed OneShot \\
\path{recorded-effect-strict-replay} & StrictExecutionReplay \code{Sound}; committed Recorded result \\
\shortstack[l]{\code{recorded-unknown-footprint-}\\\code{conditions-external-continuation}} & ExternalContinuation \code{Conditional}; unknown inherited footprint \\
\path{outcome-without-request-is-malformed} & ProjectionReplay \code{Unsound}; lifecycle-integrity blocker \\
\path{boundary-grade} & Boundary grade prerequisites \\
\path{native-grade} & Native grade and prerequisite chain \\
\bottomrule
\end{tabularx}
\caption{Coverage intent of the language-neutral v1 conformance vectors.}
\label{tab:vector-coverage}
\end{table}

\subsection{Real-log normalization}

Validation accepts ordered JSONL.  For SQLite-backed ActiveGraph stores, an
export script uses \code{events.seq} as the ordering authority and emits one
file per run.  Source domain events that are not part of the closed effect or
evidence vocabulary remain signals and are reported by type.  The adapter does
not infer effects from every \code{*.requested} event because real domains use
names such as \code{round.requested}; only a closed list of known external
request types is classified.

This choice creates an explicit adapter-coverage premise.  Unknown footprint
fails closed once an event is classified as an effect, but an unclassified
source event is not automatically treated as an effect.  A Sound continuation
verdict therefore means ``Sound under the profile declaration that the listed
unclassified types are domain-only.''  A new source type that hides an external
interaction can invalidate that verdict.  The validator reports every such type
so the declaration can be audited rather than silently assumed.

For an external request, footprint is read from explicit side-effect metadata.
A write without an idempotency key is OneShot; an unknown tool footprint remains
Unknown.  Model and embedding requests default conservatively to OneShot.  A
response is Recorded only when the log contains an explicit response hash,
output hash, or raw output from which the validator derives a content SHA-256.
The derived hash certifies captured replay material; it is not a provider
receipt or proof of external commitment.

Every source event is assigned a contiguous normalized sequence.  Bridge
events can produce multiple evidence facts; this is why the bridge normalized
count exceeds its source count.  The source event identity is retained for the
first fact, and deterministic derived identifiers are used for additional
facts.  Duplicate or non-contiguous normalized identity remains a blocker.  The
normalizer also retains the cumulative cut after each atomic source event.
These source-boundary positions are reported separately from diagnostic cuts
between co-derived facts.

\subsection{All-cut enumeration}

For a normalized log of length $n$, the validator evaluates cuts
$0,\ldots,n$.  Prefix states are computed with one scan; environmental findings
for suffixes are accumulated in reverse.  It reports both (i) every normalized
position, useful for checking the executable transformation, and (ii) only
source-event boundaries, which are the physically meaningful fork candidates
in the source trace.  The validator reports Sound, Conditional, and Unsound
totals for ProjectionReplay, ExternalContinuation, and CounterfactualWorld,
plus the number of runs containing an unsound counterfactual cut.

Actual ActiveGraph forks require a second audit because the child trace has
lineage and store-level sequence differences.  The SQL audit resolves the
parent cut event, orders the parent prefix and child trace independently, and
compares corresponding event ID, type, actor, payload, frame, cause, and
timestamp.  In this SQLite schema, \code{events.seq} is the database-wide
autoincrement primary key, not a run-local envelope field.  It is used to order
each trace and excluded from equality because copied child events receive new
store positions.  The normalizer then assigns the contiguous per-run sequence
used by Equation~\ref{eq:event-envelope}; that derived ordinal is not a second physical field in the
fork audit.  Child run ID is also excluded.  The audit separately counts
retained external requests and unresolved boundaries using the same explicit
request/outcome vocabulary as the normalizer; unknown event types remain
unclassified rather than being inferred as effects.

\subsection{Reproducibility discipline}

Evidence entries record repository, immutable revision when available, source
path, artifact hashes, counts, parser profile, command, publication status, and
limitations.  Checked-in fixtures are hash-bound and explicitly marked
synthetic or sanitized.  Private and local runtime payloads are not copied into
the repository merely to make a test convenient.  Missing revision or source
provenance is reported as a limitation rather than replaced with a hash-only
claim of reproducibility.

\section{Validation Against Runtime Artifacts}
\label{sec:validation}

We evaluate the contracts against one public trace corpus and one public
post-oracle conformance fork.  The sealed Synthetic Players capsule contains an
ActiveGraph SQLite store.  The activegraph-bridge fixture executes an actual
bridge fork after a committed recorded fixture-oracle effect.  Both contain
synthetic experimental content; ``real'' here means actual runtime artifacts
rather than invented shape fixtures.  Neither is independent of the ActiveGraph
family, and we do not claim human behavioral validity or real-model performance.

\subsection{Synthetic Players capsule}

The public source is pinned to revision \code{82556ef9dae3}.  The compressed
engine store begins SHA-256 \code{7b7fdfe64b5b}, and the capsule checksum
ledger begins SHA-256 \code{6d80f3b19b92}.  The evidence ledger records the
complete revision and digests.
Running the public verifier reports:

\begin{quote}
CAPSULE VERIFICATION PASS --- 4,919 archived Phase 3--5 runs verified
(4,916 confirmatory + 3 legacy diagnostics).
\end{quote}

The decompressed database contains 5,540 runs and 311,756 events.  Event types
include 36,251 \code{llm.requested}, 36,227 \code{llm.responded}, and 30,397
\code{decision.parsed} records.  The difference between request and response
counts is reflected rather than repaired by the adapter.

The two run counts have different denominators.  The store contains 5,505
completed and 35 incomplete runs.  The public verifier covers 4,916
confirmatory Phase 3--5 runs plus three completed legacy diagnostics.  The
remaining 586 completed runs are outside that replay contract; the 35
incomplete runs are also outside it.  Thus $4{,}919+586+35=5{,}540$.

\subsubsection{Replay grades}

All 5,540 logs grade Observed, and zero are verified \emph{from the run log
alone}.  This does not contradict the passing capsule verifier.  The capsule
attestation is external to each run's event stream.  A grading function that
claims to depend on the log alone cannot silently import that attestation.
This is a useful negative result: sealed-artifact reproducibility and
self-describing run replayability are distinct properties.

\subsubsection{All-cut results}

Table~\ref{tab:syntheticcuts} reports every cut.  The projection total is
$311{,}756 + 5{,}540 = 317{,}296$, because each run of length $n$ has $n+1$
cuts.

\begin{table}[ht]
\centering
\small
\begin{tabular}{@{}lrrr@{}}
\toprule
Property & Sound & Conditional & Unsound \\
\midrule
ProjectionReplay & 317,296 & 0 & 0 \\
ExternalContinuation & 160,076 & 120,969 & 36,251 \\
CounterfactualWorld & 150,230 & 24,172 & 142,894 \\
\bottomrule
\end{tabular}
\caption{All-cut verdicts for 5,540 Synthetic Players logs.}
\label{tab:syntheticcuts}
\end{table}

Every projection cut is sound under the specified domain-projection
observation.  The 36,251 ExternalContinuation \code{Unsound} cuts equal the number of
open/cut-through LLM requests.  Direct sequence audit explains the equality:
36,227 requests are immediately followed by their linked response, so each
creates exactly one cut between request and response; the remaining 24 requests
are the final event of failed runs, so each also creates exactly one open cut.
The 120,969 \code{Conditional} continuation cuts also have a direct per-run
identity.  For a request-bearing run with $n_r$ events, first-request ordinal
$q_r$, and $m_r$ requests, the $q_r$ earlier cuts are \code{Sound}, each request
contributes one open-boundary \code{Unsound} cut, and the remaining
$n_r+1-q_r-m_r$ closed post-response cuts are \code{Conditional}.  Summing that
last term yields 120,969; these cuts retain at least one committed OneShot oracle
interaction and require serving recorded results without another call.

CounterfactualWorld is stricter and has a separate structural identity.  In
runs whose requests are all closed, every cut before the final response has a
committed oracle interaction in the discarded suffix; the sum of those final
response ordinals is 141,545.  In the 24 runs ending with an unmatched request,
all $n_r+1$ cuts are \code{Unsound}, adding 1,349.  Thus
$141{,}545+1{,}349=142{,}894$.  The union contains 4,923 affected runs: 4,919
have at least one response, four open runs have none, and the other 20 open runs
are already in the first set.

The 160,076 Sound ExternalContinuation cuts validate only the
no-inherited-effect case.  A direct per-run ordinal query confirms that every
one lies before the first classified external request, including all cuts in
the 617 runs with no classified request.  The Synthetic Players corpus alone
therefore does not empirically validate a post-oracle continuation,
target-environment attestation, or zero-reexecution receipt.  Generated
properties and named fixtures exercise post-oracle verdicts and typed
no-reexecution obligations; the public conformance fork below separately
exercises their runtime discharge under a fixture trust root.

The validator leaves domain and infrastructure types unclassified, including
\code{behavior.started} and \code{decision.parsed}.  It likewise leaves
\code{infra.*}, \code{round.requested}, and \code{run.completed} unclassified.  In particular,
\code{round.requested} is not misclassified as an external request merely
because its name ends in \code{requested}.  Continuation verdicts are therefore
relative to the audited profile declaration that these reported types are
domain or infrastructure signals rather than hidden external effects.

\subsubsection{Observed forks}

The SQLite run table contains 121 child runs from 41 distinct parents and no
nested child in this sample.  All forks cut at a \code{round.played} event.  The
audit compares 16,625 parent-prefix events with the beginning of their child
traces and finds zero mismatches in event ID, type, actor, payload, frame, cause,
or timestamp.  Every cut event resolves.  Run identity and \code{events.seq}
are excluded because they are expected to differ: in this store schema,
\code{seq} is the database-wide autoincrement key used for ordering, not a
run-local field.  The per-run normalized sequence is derived after export and
was therefore not an additional physical field available for comparison.

None of the 121 retained prefixes contains a classified external request, and
none contains an unresolved request.  The observed forks therefore satisfy the
current ExternalContinuation precondition for a domain-only cut family.  This
is strong evidence for cheap structural branching where the runtime actually
forked.  It is not evidence for an effect-boundary fork: the observed sample
does not contain one.

\subsection{Public post-oracle conformance fork}

The companion activegraph-bridge artifact is public in the v0.2.0 release and
on the default branch at revision \code{843824a44d48}.  The complete
\href{https://github.com/yoheinakajima/activegraph-bridge/tree/843824a44d48d816779fc0c08580ae06108fe7b6/evidence/post-oracle-fork-v1}{public post-oracle fixture and verifier}
are revision-pinned.
It records an actual bridge child fork after a committed effect classified as
OneShot footprint plus Recorded replay source.  The source run makes one
deterministic offline fixture-oracle call.  Verification and the child serve
the committed outcome from the record, and the child diverges only after a
changed tool result.

The receipt binds the parent and child run identities, cut boundary, parent log
hash, exact retained-prefix hash, child log hash before receipt emission,
inherited request and outcome identities, response hash, source and target
runtime fingerprints, and target-environment claims.  The fixture generator
directly observes the oracle-call counter before and after the child operation.
The verifier checks that process-produced counter, the receipt/log hashes, and
an HMAC-SHA256 signature under a configured trust root whose caller-issued
claims include the child, cut, prefix, and target fingerprint.
The published command reports:

\begin{quote}
POST-ORACLE FORK RECEIPT PASS --- committed oracle served from record;
0 inherited external calls.
\end{quote}

The parent JSONL digest begins \code{884e8ae34296}; the child begins
\code{747305a6774a}.  The receipt file begins \code{da5503a80b92}, and its
canonical internal receipt hash begins \code{5d36df7dc177}.  The verified receipt names inherited request
\code{evt\_008}, its recorded outcome \code{evt\_009}, one source oracle call,
and zero inherited external calls in the fork.

This exercises the previously missing implementation path for a public
conformance case.  In this deterministic offline fixture, an inherited recorded
oracle outcome crosses a fork boundary with no second observed fixture-oracle
call.  A caller-issued target-environment assertion is signature-checked and
bound to that fork under a configured trust root; the verifier does not inspect
or independently establish the asserted environment contents.  The HMAC key is
deliberately published as a fixture key.  The result does not attest a real
provider, production identity, model quality, or production environmental
fidelity.

\subsection{Validation status}

\paragraph{V1: compute grade distributions.}
Completed for the public corpus.  The capsule yields Observed for 5,540 logs
because verification is external to each run log.  The result supports
log-derived grading rather than profile-derived labels and separates a sealed
artifact's verification from evidence carried by an individual run.

\paragraph{V2: inspect real forks and their preconditions.}
Completed for the public capsule and the public conformance fixture.  All 121
observed corpus forks preserve the compared prefix fields and retain no
classified external request.  The companion fixture tests one actual bridge
fork after a committed recorded oracle call.  Its generator observes no second
fixture-oracle call; its verifier checks receipt/log consistency and a
fork-bound caller assertion under a public conformance key.  It is conformance
evidence, not a real-provider or production-attestor result.

\paragraph{V3: test production confluence conditions.}
Partially completed.  Source audit confirms that ActiveGraph imposes FIFO plus
registration order, which can mask non-confluence.  The property suite bypasses
that order.  A complete inventory of production behavior reads, writes,
emissions, triggers, and relation-order dependencies was not available, so we
cannot certify or falsify a general production non-interference contract.

\paragraph{Combined interpretation.}
The evidence supports deterministic projection, structurally correct observed
prefix copying, and one public offline post-oracle fixture in which no second
fixture-oracle call was observed and a caller environment assertion was
fork-bound.  It does not support an unrestricted claim that any historical cut
is safely executable or environmentally counterfactual.  Those stronger claims
require effect-complete evidence and a property-specific precondition.

\section{Related Work and Novelty Boundary}
\label{sec:related}

\paragraph{Reducers and event sourcing.}
Mealy machines produce output and a next state from a current state and input
\cite{mealy1955method}.  Elm's update loop maps a message and model to a new
model and command \cite{elmarchitecture}.  Chassaing's functional
event-sourcing Decider separates decision and evolution and demonstrates
composition in F\# \cite{chassaing2021decider}.  These precedents invalidate a
novelty claim for the reducer shape.

Fowler's event-sourcing account makes an ordered event sequence the basis for
rebuilding application state and temporal queries, while explicitly warning
that replayed updates to external systems can go wrong unless gateways suppress
them \cite{fowler2005eventsourcing}.  Our work does not claim to discover that
external effects complicate replay.  It supplies a property-indexed cut
assessment using request, outcome, footprint, and replay-source evidence in
archived agent traces.

\paragraph{Reactive rules, termination, and confluence.}
The closest classical precedent for our scheduler counterexamples is the active
database literature.  Aiken, Hellerstein, and Widom analyze whether production
rule sets terminate, produce a unique final database state, and produce a unique
observable action stream \cite{aiken1995static}.  Paton and D\'{i}az survey ECA
rule languages and the execution-model choices that determine active-system
behavior \cite{paton1999active}.  Consequently, ``rules can trigger rules and
be order-sensitive'' is not our novelty claim.

Newman's lemma relates termination and local confluence in abstract rewriting
systems \cite{newman1942}.  CRDTs provide confluence by construction for data
types whose concurrent updates satisfy the required convergence conditions
\cite{shapiro2011crdt}.  We neither prove a minimal general condition nor
compete with these results.  Our narrower contribution is to expose the
ordering dependency in an agent runtime, preserve a read/trigger counterexample
that defeats write-set diagnostics, and connect scheduler assumptions to replay
claims.

\paragraph{Compensation and durable execution.}
The Compensatable footprint is an application of the saga lineage, where a
long-lived transaction is decomposed and partial completion is amended by
compensating transactions \cite{garciamolina1987sagas}.  This paper adds no new
compensation theory.  It asks whether a fork's retained evidence shows that the
required compensation or reconciliation actually occurred.

Temporal replays an ordered event history, requires deterministic workflow
decisions, and reuses recorded activity results rather than repeating external
work \cite{temporal2026workflows}.  Azure Durable Task similarly uses event
sourcing and constrains orchestrator code to deterministic APIs
\cite{azure2026constraints}.  DBOS checkpoints steps, requires deterministic
workflow structure and idempotent steps, and does not re-execute a completed
checkpoint \cite{dbos2026architecture}.  Restate journals side-effecting
operations and results so completed steps can be skipped on recovery
\cite{restate2026concepts}.  These systems are strong industrial precedents for
``deterministic replay after effect results are fixed.''  Our delta is not the
mechanism; it is a per-cut, property-relative license that distinguishes
projection, zero-call execution, continuation in a supplied environment, and a
counterfactual-world observation.

LangGraph exposes checkpoint replay and fork operations.  Its documentation
states that nodes before a checkpoint are skipped while later nodes re-execute,
including LLM and API calls \cite{langgraph2026timetravel}.  This is precisely
why a product-level ``time travel'' operation needs an effect-boundary contract:
state branching and external re-execution are separate questions.

\paragraph{Agent calculi and mechanized semantics.}
$\lambda_A$ gives a typed lambda calculus for agent composition with oracle
calls, bounded fixpoints, probabilistic choice, and mutable environments, with
machine-checked type-safety and bounded-termination results
\cite{liu2026lambdaa}.  We study runtime event logs, schedule-sensitive
settlement, and effect-boundary cuts rather than composition-time typing.

LLMbda models agent conversations, fork and clear operations, code generation,
and labeled information flow.  Its central result is machine-checked
termination-insensitive probabilistic noninterference
\cite{garby2026llmbda}.  Our work does not claim noninterference or
prompt-injection safety.  Schlapbach formalizes Schema-Guided Dialogue and MCP
in a process calculus and studies protocol-level bisimilarity and expressivity
\cite{schlapbach2026protocols}; we instead examine effect evidence within a
runtime trace.

\paragraph{Runtime substrates and governance.}
Agent libOS supplies a capability-controlled runtime with process identity,
memory, skills, checkpoints, restore, fork, and commit
\cite{zhang2026libos}.  It demonstrates operational mechanisms and a safety
benchmark.  Our question is which evidentiary conditions make a particular
event-log cut sound for a named observation.

McCann's Effect-Transparent Governance formalizes handler-mediated effect
governance in Rocq and proves semantic preservation, expressive minimality, and
decidability boundaries \cite{mccann2026governance}.  We concede minimality and
use an effect-transparent boundary as an executable assumption.  AgentSpec
defines trigger, predicate, and enforcement rules for agent safety
\cite{wang2026agentspec}.  Its ICSE 2026 contribution is runtime policy
enforcement, not replay licensing.

\paragraph{ActiveGraph.}
ActiveGraph makes the event log authoritative and presents deterministic
replay, cheap forks, and lineage as architectural properties
\cite{nakajima2026log}.  This paper is explicitly a follow-on.  It preserves
the fixed-log projection result, distinguishes canonical ordering from schedule
independence, and reframes forking as a property-indexed semantic contract.
The public implementation and downstream artifacts supply the validation
substrate, while their shared lineage limits external generalization.

\paragraph{Bounded novelty claim.}
After accounting for active rules, sagas, durable workflows, agent calculi, and
checkpoint products, the remaining claim is intentionally small.  We are not
aware of prior work that jointly preserves runtime-order counterexamples,
partitions effect evidence along footprint/source/lifecycle dimensions,
assesses every historical cut under multiple fork properties, and checks those
contracts against both archived traces and observed forks from a deployed
event-sourced agent runtime.  The empirical corpus is ActiveGraph-family and
the search is not systematic; both limits are explicit.

\section{Discussion}
\label{sec:discussion}

\subsection{``The log replayed'' is not one assertion}

Figure~\ref{fig:replay-family} states the hierarchy early because it is the
paper's organizing claim.  None of its relations automatically implies all the
others.  The Synthetic Players capsule passes an external byte-exact verifier,
but its individual logs contain no verification attestation, so they remain
Observed under a log-alone grade.  Every prefix projects, but cuts through
requests fail execution continuation.  The public bridge fixture then shows a
different predicate again: one committed recorded oracle result is inherited
across a fork with zero new calls under a fork-bound caller assertion.  These
are not contradictions; they are different predicates.

\subsection{Cheap fork is semantic before it is fast}

Sharing an immutable prefix is computationally cheap.  The harder question is
what the child is allowed to do next.  A runtime that advertises \code{fork(k)}
should specify at least:

\begin{itemize}[leftmargin=*]
  \item the observation under which parent prefix and child are equivalent;
  \item the ordering authority for the prefix;
  \item whether a cut may split a request from its outcome;
  \item how recorded results are served without re-execution;
  \item which retained external effects are inherited;
  \item whether the child receives an isolated/snapshotted environment or the
  already-mutated source environment;
  \item the treatment of external effects in a retrospectively discarded
  suffix.
\end{itemize}

Without these clauses, ``cheap'' describes storage cost while leaving semantic
cost undefined.  With them, the implementation can refuse unsafe cuts, require
an idempotency key, request compensation, select a snapshot, or downgrade the
claim to ProjectionReplay.

A proposed copyable API contract is:

\begin{equation}
\mathsf{fork}(L,k,P) \rightarrow
\mathsf{Ok}(child,receipt)\;|\;
\mathsf{Refuse}(reason)\;|\;
\mathsf{Conditional}(obligation).
\end{equation}

The executable assessment represents a Conditional payload with the closed
union

\begin{align}
\mathsf{Obligation} ::= {}&
  \mathsf{ServeRecordedResultWithoutReexecution}(effectId) \\
  \mid{}& \mathsf{ResolveUnknownInheritedFootprint}(effectId) \\
  \mid{}& \mathsf{ReconcileDiscardedEffect}(effectId) \\
  \mid{}& \mathsf{ApplyCompensation}(effectId) \\
  \mid{}& \mathsf{ReconcileFailedEffect}(effectId).
\end{align}

The proposed receipt binds the parent, cut, observation $P$, prefix hash,
environment premise, inherited effect identities, and the typed obligation
set.  The F\# assessor computes findings and this obligation set.  The companion
public activegraph-bridge v0.2 implementation emits a hash-bound
receipt and signature-checks a fork-bound, caller-issued target-environment
claim under a configured trust root; it does not validate the claim's contents.
The public post-oracle fixture exercises that path.  This is a conformance
mechanism, not a production identity, provider attestation, or environmental
proof.  In particular, a cut through Requested is refused unless an explicit
recovery policy changes the requested property.

\subsection{Deterministic ordering can conceal non-confluence}

FIFO and registration order are useful engineering contracts.  They make one
execution repeatable and improve debuggability.  But they can turn an accidental
ordering dependency into a stable behavior, making ordinary replay tests pass
while schedule independence remains false.  This matters when a runtime later
parallelizes handlers, changes relation storage, distributes a queue, or moves
from one database backend to another.

The appropriate response need not be to demand full confluence.  Some systems
may intentionally make order semantic.  The contract should then name the
canonical order, preserve it in the log, and reject backends that cannot supply
it.  Systems that want schedule independence need a stronger behavior
non-interference discipline.  The current work distinguishes the choices but
does not prescribe one universally.

\subsection{The log should carry its own licenses}

The gap between capsule verification and log-alone grade suggests a design
principle: evidence needed to authorize later replay or fork should be retained
with the run or its signed sidecar.  A minimum receipt should include canonical
sequence, run identity, parent and cut lineage, effect request identity,
footprint annotation, outcome linkage, response content hash, commitment or
failure evidence, reconstruction method, and verification result.

This does not require every event to be trusted merely because it is present.
The event type must be recognized, the producer or signature may need
authentication, and cross-event integrity must be checked.  The present
F\# artifact enforces structural integrity and exact recognized types; the
companion bridge adds an HMAC-signed sidecar under a configured trust root.
Production use still needs an operationally protected key, authoritative
environment identity, rotation and revocation policy, and a freshness contract.
A future asymmetric or transparency-backed evidence capsule could make that
trust portable across implementations and organizations.

\subsection{The graph may be a projection, but the scheduler remains real}

Replayed domain state can be treated as a projection.  That reduction still
omits two operational facts.  Once multiple
behaviors react, event and activation order become part of the operational
semantics unless a confluence condition removes them.  Once effects leave the
process, the external world becomes a second state that the event projection
can describe but not reverse.

Thus reducing the runtime to a pure fold is useful but incomplete.  The deeper
object is an event-producing transition system plus a scheduler, an effect
interpreter, and evidence linking interpreter outcomes back to the log.  The
graph may be derived; ordering and effect contracts remain non-eliminable
obligations of the operational semantics.

\subsection{Operational recommendations}

The results suggest concrete contracts for event-sourced agent runtimes:

\begin{enumerate}[leftmargin=*]
  \item Treat production order as an explicit scheduler policy and test at
  least event-order and activation-order perturbations.
  \item Record behavior read, write, emission, and trigger dependencies if
  schedule independence is a goal; a write set alone is insufficient.
  \item Separate replay source from external footprint.  ``Cached'' does not
  mean ``never happened'' or ``safe to repeat.''
  \item Make request and terminal outcome a one-to-one integrity boundary.
  Refuse a cut through Requested or Unknown without an owner-selected recovery
  policy.
  \item Give every fork API a property parameter or an equivalently explicit
  mode.  Return a receipt, refusal, or conditional obligation rather than a
  single unqualified \code{safe} Boolean.
  \item Preserve verification and reconstruction evidence inside a portable,
  hash-bound run capsule.
  \item Distinguish actual observed forks from every-cut counterfactual tests.
  The former show where users forked; the latter reveal untested hazards.
\end{enumerate}

These recommendations are compatible with F\#, C\#, Python, Rust, or
TypeScript.  The conformance vectors intentionally encode events and expected
verdicts in language-neutral JSON so another implementation can challenge the
semantics.

\section{Limitations and Threats to Validity}
\label{sec:limitations}

\paragraph{No mechanized proof.}
The propositions and contracts are encoded as executable properties and checked over finite
generators and artifacts.  A passing FsCheck property is not a proof over all
behaviors, schedulers, or logs.  Future work may mechanize selected definitions
after the operational contracts stabilize.

\paragraph{No minimal confluence condition.}
We falsify general confluence and declared-write disjointness.  We check a
restricted isolated-writer family.  We do not establish a necessary or
sufficient general condition over reads, writes, emissions, triggers, and state
domains.  Calling the current restricted condition ``the'' restoring condition
would overstate the result.

\paragraph{Limited scheduler space.}
The executable suite uses four deterministic scheduler extremes.  They expose
the preserved counterexamples but do not enumerate every interleaving for an
arbitrary behavior set.  Relation ordering is discussed from source audit but
not modeled as a separate production activation dimension.  Seeded randomized
and bounded exhaustive interleaving exploration remain future checks.

\paragraph{Kernel/production read timing.}
The executable kernel snapshots state for all activations enabled by one
trigger.  The pinned ActiveGraph source computes the match set once but rebuilds
each handler view after prior handlers may have emitted and synchronously
projected events.  The source audit identifies this additional ordering seam;
the checked kernel properties do not model it, and no production behavior
inventory establishes whether it changes an observed run.

\paragraph{Adapter assumptions.}
Real-log classification relies on explicit event names and payload fields.  The
adapter is deliberately conservative, but mistakes remain possible.  Model
calls are treated as OneShot for cost/oracle semantics; another property might
classify a read-only deterministic local model differently.  A derived content
hash proves captured bytes, not provider authenticity, commitment, or
idempotence.

The generic F\# effect descriptor and legacy corpus adapters have no per-effect
observable-set label.  Their CounterfactualWorld results therefore use the
selected profile's entire recognized external surface as
$\Omega_{profile}$.  The public bridge v0.2 receipt records
observable tags, but we do not retroactively impute them to older corpus events.
Narrower corpus claims require adapter-supplied tags and new conformance
vectors.

Unknown footprint fails closed only after the adapter recognizes an event as an
effect.  Unclassified source events remain reported signals.  The all-cut Sound
counts therefore depend on an audited profile declaration that those event
types do not hide external interactions.  A newly introduced effect type can
make an old verdict false until the adapter is updated.

\paragraph{Assessment snapshot and delayed hazards.}
Every verdict is relative to the finite log snapshot presented to the assessor.
An asynchronous failure, compensation result, webhook timeout, or other hazard
may be appended after that assessment and downgrade a previously clean prefix.
The artifact does not model an observation-delay distribution or prove that a
source log is closed.  Production admission therefore requires a receipt bound
to a source high-water mark plus an explicit closure, quiescence, or freshness
policy and a recheck immediately before continuation.  A time window alone is
an operational assumption, not proof that no later hazard exists.

\paragraph{Partial failure and compensation scope.}
Failed remote operations may have partially committed before reporting an
error.  The assessor conservatively conditions or blocks discarded failures by
footprint, but it cannot inspect the remote system.  Likewise, successful
compensation does not erase a temporal exposure window or third-party secondary
effects.  A caller whose $\Omega$ includes those histories needs stronger
reconciliation evidence than the current effect descriptor records.

\paragraph{Whole-log malformedness policy.}
The current assessor treats structural malformedness anywhere in the source log
as a blocker for every cut, including projection.  This is fail-closed and
appropriate for the evidence study, but it is stricter than a theorem about an
independently authenticated well-formed prefix.  The projection suffix identity is
therefore stated only for well-formed ordered logs.

\paragraph{Synthetic content.}
The Synthetic Players capsule contains archived experimental runs, but the
present analysis concerns runtime traces rather than human ground truth.  The
public bridge fixture uses a deterministic offline fixture and explicitly makes
no real-model performance claim.

\paragraph{Fork coverage.}
The 121 observed forks are useful positive evidence, but all cut at
\code{round.played} and retain no classified external request.  No observed
fork in the public capsule crosses a recorded oracle or irreversible effect
boundary, and all 160,076 Sound corpus ExternalContinuation cuts precede the
first classified effect.  The companion public conformance fixture now covers
one actual bridge fork after a committed recorded oracle effect.  Its generator
observes no second fixture-oracle call, while its verifier checks receipt/log
consistency and a signed, fork-bound caller assertion under a configured trust
root.  This exercises the previously absent public no-reexecution mechanism,
but only for one deterministic offline oracle, one child, a process-produced
counter, and a deliberately published fixture key.  The verifier does not
inspect the asserted environment.  The result does not establish real-provider
behavior, production attestor identity, environmental fidelity, or prevalence
across production forks.

\paragraph{Production behavior inventory.}
The ActiveGraph source establishes FIFO and registration-order scheduling, but
we did not recover a complete deployed behavior dependency inventory.  We
therefore cannot say whether production behaviors satisfy a future
non-interference condition.  Static source inspection of the scheduler is not
production confluence validation.

\paragraph{No external runtime corpus.}
The public corpus, observed forks, and bridge implementation all descend from
ActiveGraph.  The language-neutral vectors support other implementations, but
no LangGraph, Temporal, or other independent runtime trace has yet been
normalized.  Generalization beyond the source family is therefore conceptual
rather than empirically demonstrated.

\paragraph{Baseline conformance-vector coverage.}
The seven language-neutral vectors cover the boundary cases listed in
Table~\ref{tab:vector-coverage}; they do not enumerate the full verdict matrix.
Only the unknown-inherited-footprint obligation has a dedicated Conditional
vector.  Dedicated portable vectors for no-reexecution, discarded-effect
reconciliation, compensation, and failed-effect reconciliation remain future
work, although the F\# suite exercises those obligations directly.

\paragraph{No authorization or information-flow result.}
Effect descriptors classify replay and environmental conditions; they do not
grant capabilities or prove that a caller is authorized.  The artifact also
does not track secrecy labels or prove noninterference.  Those topics are
covered by distinct systems and calculi and remain explicit non-goals.

\paragraph{No performance benchmark.}
The all-cut implementation is efficient enough for the reported corpora, but
we make no throughput, storage, or latency claim.  ``Cheap'' in this paper is
analyzed semantically.  A separate benchmark would be needed for production
performance.

\section{Conclusion}
\label{sec:conclusion}

Deterministic projection of a fixed log is valuable, but it is the beginning of
an agent-runtime replay contract rather than the end.  Reactive settlement can
diverge, a stable production order can mask non-confluence, and disjoint
declared writes do not exclude schedule-sensitive reads or trigger
interference.  A fork can preserve its trace prefix while failing to reproduce
execution or the external world.

The executable contract makes those distinctions checkable.  Effect
footprint, replay source, and lifecycle answer separate questions.  Fork safety
is indexed by an observation: projection, strict execution replay, continuation
in a prefix-reconstructed environment, or counterfactual-world equivalence.
Replay grades summarize retained evidence, while cut-level assessment decides
whether a particular branch meets the property-specific preconditions.

The runtime artifacts support both positive and negative conclusions.  Across
5,540 public capsule logs, every one of 317,296 projected prefixes is sound
under the stated projection observation.  Across 121 observed forks, 16,625
shared-prefix events show zero structural mismatches modulo expected lineage
fields, and the observed corpus cuts avoid external requests.  Yet 36,251 hypothetical
cuts split model requests and are unsound for continuation, while thousands of
discarded oracle interactions prevent an unchanged-world claim.  A separate
public offline bridge fixture crosses one committed recorded oracle boundary
and directly observes no second fixture-oracle call; its verifier checks
receipt/log consistency and a fork-bound caller assertion under a public
conformance key.

The practical conclusion is concise: replay is a family of assertions, and a
fork is sound only relative to one of them.  Event-sourced agent runtimes should
record the evidence that licenses each assertion, expose ordering assumptions,
and refuse to turn a cheap prefix copy into an unqualified promise about
execution or the world.

\bibliographystyle{plain}
\bibliography{references}

\end{document}